\documentclass[twocolumn,twoside,slac_two]{revtex4}
\usepackage{graphicx}
\usepackage{fancyhdr}
\usepackage{graphics}
\usepackage{epstopdf}
\usepackage{textpos}
\begin{document}

\title{\centering $D^0$-mixing/CPV and $D$ decays}
\author{
\centering
\begin{center}
E. Won
\end{center}}
\affiliation{\centering 
Physics Dept. Korea University Seoul 136-713, Republic of Korea}
\begin{abstract}
 We review basic phenomenology on $D^0$ mixing/$CP$ violation 
and recent experimental results on them. 
$D^0$ mixing is established by combining results from
multiple experiments but no $CP$ violation in the 
charm sector has been seen. $D^0$ mixing from a single 
experiment will clarify the size of the mixing,
and observation of $CP$ violation in charm
decays at the present level of experimental sensitivity
would be clear signal of new physics beyond the standard
model.
\end{abstract}

\maketitle
\thispagestyle{fancy}


\section{Introduction}

 Mixing of the strangeness flavor in the kaon system has been observed 
more than 50 years ago~\cite{ref:kaon_osc}
and $CP$ violation in the kaon system 
has also been well studied indicating, that there is no new physics (NP)
beyond the standard model (SM) in the kaon system~\cite{ref:bigi_cp}. 
Also, the oscillation
of the $b$-quark flavor, in both $B_d$ and $B_s$ meson systems, 
has been established firmly by $B$-factories and by Tevatron and
has been a leading topic in the flavor community of high energy 
physics~\cite{ref:bmeson_cp}. $CP$ violation effects in 
the $B$ meson system have also
been extensively carried out over last 10 years and we still see
no strong evidence of $CP$ violation beyond the SM~\cite{ref:pdg2010}. 

The study
of mixing and $CP$ violation of the $D$ meson is crucial since it
involves only the up-type quark, which has never been studied before. 
Recently, mixing of $D^0$  
meson has been seen by combining multiple experiments~\cite{ref:hfag}. 
However,
in general the intepretation is rather difficult since the
effect of the final state interactions are not really calculable
in the SM~\cite{ref:falk}. Becuase of this, the goal 
of search for 
mixing in the $D$ meson system is rather to probe NP, not to 
constrain Cabibbo-Kobayashi-Maskawa (CKM) matrix.
In the SM, indirect $CP$ violation in the charm system is
expected to be as small as $\mathcal{O}(10^{-4})$ and 
universal between $CP$ eigenstates~\cite{ref:grossman}. 
On the other hand, the direct $CP$ violation can be larger in
SM, depending on the final state of the $D$ decay of interest.
In the following sections, we briefly discuss basic phenomenology
of $D$ meson mixing and $CP$ violation and the corresponding experimental 
results. 

\section{$D^0$ Meson Mixing}

\subsection{Mixing Parameters}

 In order to describe the time development of neutral $D$-meson
system, one starts with writing Schr\"odinger equation for a
column vector that is composed of $D^0$ and $\overline{D}^0$ states:
\begin{eqnarray} 
i \frac{d}{dt} 
\left(
\begin{array}{c}
D^0(t) \\
\overline{D}^0(t) 
\end{array}
\right)
=
\Bigg[
\mathbf{M} - \frac{i}{2} \mathbf{\Gamma}
\Bigg]
\left(
\begin{array}{c}
D^0(t) \\
\overline{D}^0(t) 
\end{array}
\right),
\label{eq:sch}
\end{eqnarray} 
where $\mathbf{M}$ and $\mathbf{\Gamma}$ are 2$\times$2 matrices
that are associated with 
$(D^0,\overline{D}^0)$
$\leftrightarrow$
$(D^0,\overline{D}^0)$
transitions via off-shell (dispersive), and on-shell (absorptive)
intermediate states,
respectively~\cite{ref:pdg2010}. Diagonal elements of the
effective Hamiltonian $\mathbf{H}$ $\equiv$ $\mathbf{M} - \frac{i}{2}
\mathbf{\Gamma}$ are associated with the flavor-conserving transitions,
while off-diagonal elements are associated with flavor-changing 
transitions such as $D^0 \leftrightarrow \overline{D}^0$. The eigenstates
of the above Schr\"odinger equation are parameterized as
\begin{eqnarray}
|D_1 \rangle &\propto& p\sqrt{1-z}|D^0\rangle +
q\sqrt{1+z}|\overline{D}^0 \rangle
\nonumber \\
|D_2 \rangle &\propto& p\sqrt{1+z}|D^0\rangle -
q\sqrt{1-z}|\overline{D}^0 \rangle
\end{eqnarray}
using the notation introduced in Ref.~\cite{ref:pdg2010}. 
Parameters
$p,q$ and $z$ are complex-valued ones that relate flavor to
mass eigenstates for the $D$-meson system.
The normalized mass difference and the width difference are parameterized
as $x$ and $y$:
\begin{eqnarray}
x &\equiv& (m_1 - m_2)/\Gamma,
\nonumber \\
y &\equiv& (\Gamma_1 -\Gamma_2)/2\Gamma,
\end{eqnarray}
where $m_i$ and $\Gamma_i$ are mass and decay rate values of the eigenstate
$|D_i\rangle$ ($i=1,2$). $\Gamma$ is the average of two $\Gamma_i$s.
The parameters $x$ and $y$ are commonly called 
mixing parameters in $D$-meson decays and are experimentally measurable.
SM calculations based on box diagrams alone give $x \sim 10^{-5}$ and
$y \sim 10^{-7}$~\cite{ref:falk}, but are increased due to the long-distance
effects. The parameter $y$ is dominated by long-distance effects and is
generally considered to be insensitive to new physics. Therefore,
$x \gg y$ would point to NP phenomena~\cite{ref:wilkinson}. 

 In the case when $D^0$ decays to non-$CP$ eigenstates, for example
$D^0 \rightarrow K^\mp \pi^\pm$, one can form four different combinations
of amplitudes as 
$\overline{A}_f=\langle K^+\pi^-|\mathcal{H}|\overline{D}^0\rangle$,
$A_{\overline{f}}=\langle K^-\pi^+|\mathcal{H}|{D}^0\rangle$,
${A}_f=\langle K^-\pi^+|\mathcal{H}|\overline{D}^0\rangle$,
and
$\overline{A}_{\overline{f}}=\langle K^+\pi^-|\mathcal{H}|{D}^0\rangle$
where the first two are called ``right-sign'' decay amplitudes as
they are Cabibbo favored (CF), and latter two are ``wrong-sign'' decay
amplitudes as they are doubly Cabibbo-suppressed decays (DCSD) or
proceed through mixing. Conventionally one normalizes the 
wrong-sign decay distributions
to the integrated rate of right-sign decays to define $r(t)$ and $\overline{r}
(t)$:
\begin{eqnarray} 
r(t) &\equiv& \frac{|\langle f|\mathcal{H}|D^0(t)\rangle|^2}{|\overline{A}_f|^2}
=
\Bigg|
\frac{q}{p}
\Bigg|^2
\Big|
g_+ (t) \lambda^{-1}_f + g_-(t)
\Big|^2,
\nonumber \\
\overline{r}(t) &\equiv& \frac{|\langle \overline{f}|
\mathcal{H}|\overline{D}^0(t)\rangle|^2}{|{A}_{\overline{f}}|^2}
=
\Bigg|
\frac{p}{q}
\Bigg|^2
\Big|
g_+ (t) \lambda_{\overline{f}} + g_-(t)
\Big|^2 
\end{eqnarray} 
where $\lambda_f \equiv q\overline{A}_f/p A_f$,
$\lambda_{\overline{f}} \equiv q\overline{A}_{\overline{f}}/p A_{\overline{f}}$,
and
$g_\pm(t) = \frac{1}{2}(e^{-iz_1 t} \pm e^{-iz_2 t})$ with
$z_{1,2} = {\omega_{1,2}}/{\Gamma}$. $\omega_{1,2}$ are the
eigenvalues of Eq.~(\ref{eq:sch}).

\begin{figure}
\includegraphics[width=65mm]{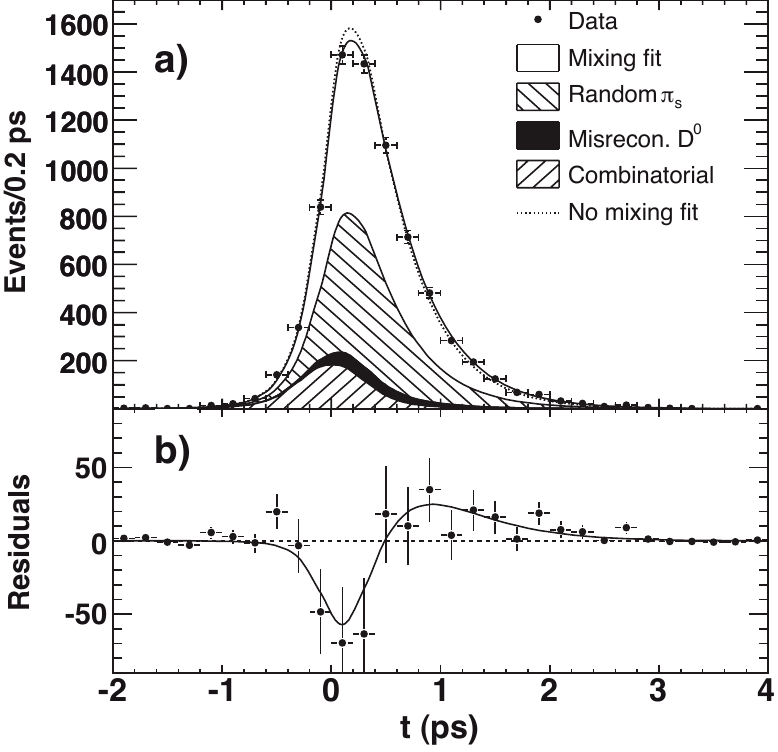}
\caption{Projections of the proper-time distribution of combined $D^0$ 
and $\overline{D}^0$ wrong-sign candidates (points) and fit result of
$D^0 \rightarrow K^+ \pi^-$ in (a). Fit results with and without
mixing hypotheses are included as solid and dashed curves, respectively.
(b) The points represent the difference between the data and the no-mixing 
fit. The solid curve shows the difference between fits with and 
without mixing. BaBar collaboration, Ref.~\cite{ref:babar_ws}.
\label{fig:babar_ws}
}
\end{figure}

\subsection{Semi-leptonic Decays}

 Let us consider the final state of $f=K^+ \ell^- \overline{\nu}_\ell$. 
In this case, $A_f = \overline{A}_{\overline{f}}$ = 0 within the SM.
The final state $f$ is only accessible through mixing and one can
obtain
\begin{eqnarray}
r(t) &=& |g_-(t)|^2 
\Big|
\frac{q}{p}
\Big|^2 
\approx \frac{e^{-t}}{4} (x^2+y^2)t^2 
\Big|
\frac{q}{p}
\Big|^2 
\nonumber \\
\overline{r}(t) &=& |g_-(t)|^2 
\Big|
\frac{p}{q}
\Big|^2 
\approx \frac{e^{-t}}{4} (x^2+y^2)t^2 
\Big|
\frac{p}{q}
\Big|^2.
\label{eq:semi}
\end{eqnarray}
Note that in the 
SM, $CP$ violation in charm mixing is small and $|q/p| \approx 1$
is satisfied. Also, in the limit of $CP$ conservation, $r(t)=\overline{r}(t)$.
From Eq.~(\ref{eq:semi}), one can compute the time-integrated mixing
rate relative to the time-integrated right-sign rate for semi-leptonic
decays, $R_M$, as
\begin{eqnarray}
R_M = \int^\infty_0 r(t)~dt = \frac{1}{2}(x^2+y^2),
\end{eqnarray}
which is a circle in $x-y$ plane. The present world-average value of $R_M$
found by the heavy flavor averaging group~\cite{ref:hfag} 
is $R_M = (1.30 \pm 2.69)\times 10^{-4}$. The most sensitive 
estimation of $R_M$ is carried out by the Belle 
collaboration~\cite{ref:belle_semi}. Using the decay mode of
$D^{*+} \rightarrow D^0 \pi^+_s$ when 
$D^0 \rightarrow K^{(*)-} \ell^+
\nu$ (right-sign) or 
$D^0 \rightarrow \overline{D}^0 \rightarrow K^{(*)+} \ell^- \overline{\nu}$
(wrong-sign), one can infer the flavor of the $D$ meson at the
production by identifyig the charge of the slow pion ($\pi^+_s$).
By counting the yields of right-sign and wrong-sign decays one
extract $R_M = (1.3\pm 2.2 \pm 2.0)\times 10^{-4}$ or
$R_M < 6.1\times 10^{-4}$ at 90\% confidence level (C.L.)~\cite{ref:belle_semi}. 

\subsection{Wrong-Sign Decays}
For the final state $f=K^+ \pi^-$, one parameterizes the ratio
of decay amplitude as
\begin{eqnarray}
\frac{A_f}{\overline{A}_f} = - \sqrt{R_D} e^{-i \delta_f},~~~\textrm{with}
~
\Bigg|
\frac{A_f}{\overline{A}_f}
\Bigg|
\sim \mathcal{O}(\tan^2{\theta_c})
\end{eqnarray}
where $R_D$ is the decay rate ratio of DCSD to CF modes, 
and $\delta_f$ is the strong phase difference between them. 
If we introduce three $CP$ violating, real-valued parameters $A_M$,
$A_D$, and $\phi$,
in this wrong-sign decays, to leading order ($A_D,A_M \ll 1$), one
can write
\begin{eqnarray}
r(t) &=& e^{-t}
\Big[
R_D (1+A_D) + \sqrt{R_D(1+A_M)(1+A_D)}
\nonumber \\
&\times& y^\prime_- t 
+\frac{1}{2}
(1+A_M)R_Mt^2
\Big]
\nonumber \\
\overline{r}(t) &=& e^{-t}
\Big[
R_D (1-A_D) + \sqrt{R_D(1-A_M)(1-A_D)}
\nonumber \\
&\times& y^\prime_+ t 
+\frac{1}{2}
(1-A_M)R_Mt^2
\Big]
\end{eqnarray}
where $y^\prime_\pm \equiv y^\prime \cos{\theta} \pm x^\prime
\sin{\phi}$, 
$x^\prime = x\cos{\delta_{K\pi}} 
+ y \sin{\delta_{K\pi}}$, and
$y^\prime = y\cos{\delta_{K\pi}} 
- x \sin{\delta_{K\pi}}$. Note that $\delta_{K\pi}$ is the relative
strong phase between final state $K$ and $\pi$, and therefore extraction
of a different set of mixing parameters $x^\prime$ and $y^\prime$ 
in this final state requires the value of $\delta_{K\pi}$. 
An interference effect in the decay chain
provides useful sensitivity to $\delta_{K\pi}$ and is discussed later.
The BaBar collaboration looks at the wrong-sign decay $D^0 \rightarrow
K^+ \pi^-$ and fits the proper time distribution as shown in
Fig.~\ref{fig:babar_ws}. From this, one extract the mixing parameters
as $x^{\prime 2} = (-0.22\pm0.33\pm0.21)\times10^{-3}$ and
$y^{\prime} = (9.7\pm4.4\pm3.1)\times10^{-3}$~\cite{ref:babar_ws}.
clearly the data prefer the mixing hypothesis.

\subsection{Determination of Strong Phase}

 The decay of the 
quantum-coherent $C=-1$ state, $\psi(3770) \rightarrow D^0 \overline{D}^0$
provides time-integrated sensitivity to the strong phase. The neutral
charm meson in the CLEO-c program does not travel far enough for time-dependent
study. Using the relations
\begin{eqnarray}
\cos{\delta_{K\pi}} =
\frac{|A(D_+ \rightarrow K^-\pi^+)|^2 - |A(D_- \rightarrow K^-\pi^+)|^2}{
2\sqrt{R_D}|A(D^0 \rightarrow K^-\pi^+)|^2
}
\nonumber
\end{eqnarray}
where $D_\pm$ denotes a $CP$-even or -odd eigenstate, one can have 
experimental access to $\delta_{K\pi}$.
CLEO-c uses $CP$-tagged
decays to obtain 
$\cos{\delta_{K\pi}} = 1.10\pm0.35\pm0.07$ or
$\delta_{K\pi} = \Big(22^{+11+~9}_{-12-11}\Big)^\circ$~\cite{ref:cleo_delta}.

\begin{figure}
\includegraphics[width=65mm]{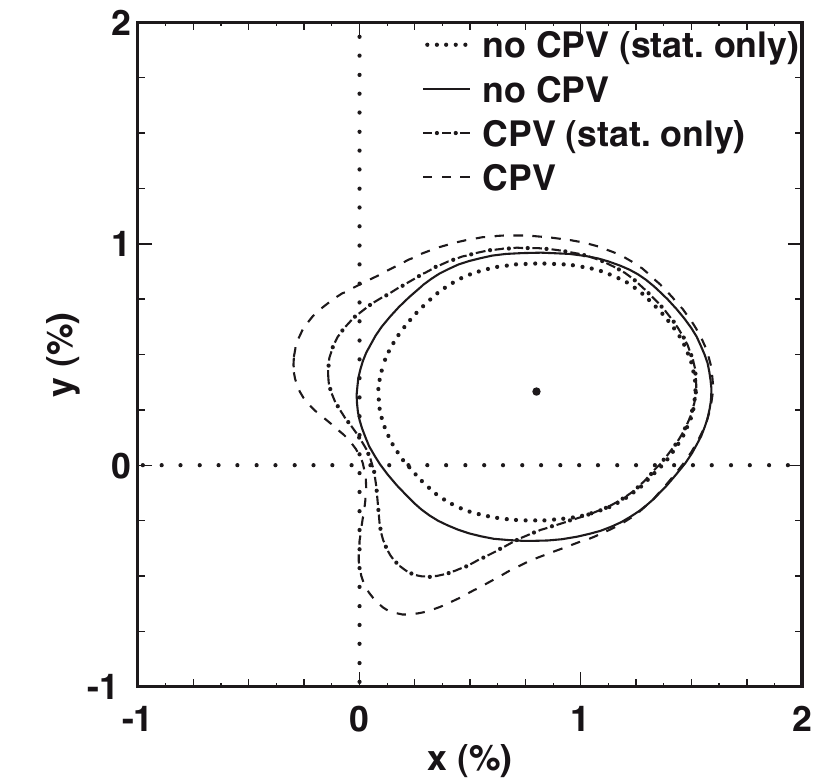}
\caption{
Belle collaboration
95\% C.L. contours for ($x,y$): dotted (solid) corresponds to 
statistical (statistical and systematic) contour for no $CP$
violation, and 
dash-dotted (dashed) corresponds to statistical 
(statistical and systematic) contour for the $CP$-allowed
case~\cite{ref:belle_dalitz}. 
The point is the best-fit result for no $CP$ violation.
\label{fig:belle_dalitz}
}
\end{figure}

\subsection{Decays to $CP$ Eigenstates}
When $D^0$ mesons decay to $CP$ eigenstates, for example $f=K^+K^-$,
there is no distinction between $f$ and $\overline{f}$, and therefore
$A_f = A_{\overline{f}}$ and
$\overline{A}_{\overline{f}} = \overline{A}_{{f}}$.
In this case, we define $y_{CP}$ and $A_\Gamma$ in order to probe
the amount of indirect $CP$ violation as
\begin{eqnarray}
y_{CP} &\equiv& \frac{\Gamma_{\overline{D}^0 \rightarrow K^+K^-}
+\Gamma_{D^0 \rightarrow K^+K^-}}{2\Gamma} -1 
\nonumber \\
&\approx& y \cos{\theta} -\frac{1}{2}A_M x \sin{\theta},
\nonumber \\
A_\Gamma &\equiv& \frac{\Gamma_{\overline{D}^0 \rightarrow K^+K^-}
-\Gamma_{D^0 \rightarrow K^+K^-}}{2\Gamma}  
\nonumber \\
&\approx& \frac{1}{2}A_M y \cos{\theta} - x \sin{\theta}.
\end{eqnarray}
Note that in the limit of $CP$ conservation, we expect that
$y_{CP}=1$ and $A_\Gamma=0$ and therefore they are parameters
that probe $CP$ violation phenomena in $D^0$-meson decays.
Substantial work on the time-integrated $CP$ asymmetries in decays
to $CP$ eigenstates are carried out and so far all are consistent 
with no $CP$-violation at $\sim\mathcal{O}(1)$\% level. The
BaBar collaboration has studied $D^0 \rightarrow K^+K^-/\pi^+ \pi^-$
decays to extract a value for $y_{CP}$
and extracted $y_{CP}$ value. Experimentally, $y_{CP} =
\langle \tau_{K\pi}\rangle / \langle \tau_{hh}\rangle -1$, where
$\langle \tau_{hh} \rangle = 
(\tau^{D^{0}}_{hh}+\tau^{\overline{D}^{0}}_{hh})/2$ and
is measured to be
$y_{CP} = (1.16\pm 0.22\pm 0.18)\%)$~\cite{ref:babar_ycp} 
consistent with 
no $CP$ violation hypothesis. The Belle collaboration has measured
the decay-rate asymmetry for the $CP$-even final states $A_\Gamma$
by separately determining the apparent lifetimes of $D^0$ and 
$\overline{D}^0$ in decays to the $CP$ eigenstates as
$A_\Gamma=(0.01\pm0.30\pm0.15)\%$~\cite{ref:belle_mix}. 
Again, the result is consistent
with the assumption of no $CP$ violation in the charm sector.

\subsection{Dalitz Analysis}

 Dalitz analysis is an invaluable technique exploited in many
charm analyses. In general, for a three-body decay $D\rightarrow
A+B+C$, one can fully describe the kinematics of such decay using
two parameters $m^2_{AB} \equiv (p_A + p_B)^2$ and 
$m^2_{BC} \equiv (p_B + p_C)^2$. They are extremely useful since
they are Lorentz invariant, the phase space of the decay is flat,
and because of that, possible two-body resonances can be clearly
seen. This technique has been used in light meson spectroscopy,
CKM angle $\phi_3(\gamma)$ measurements, and mixing/$CP$ 
violation studies.
Note that this technique can easily be extended to four and higher number
body
decays. In the decay of $D^0 \rightarrow K^0_S \pi^+ \pi^-$, there
are many quasi two-body intermediate states such as
$D^0\rightarrow K^{*-}\pi^+$ (CF),
$D^0\rightarrow K^{*+}\pi^-$ (DCSD),
and
$D^0\rightarrow \rho^{0}K^0_S$ ($CP$ eigenstate). Therefore, one
form a total amplitude as
\begin{eqnarray}
\mathcal{A} (m^2_-,m^2_+) = \sum_r a_r e^{i\phi_r} \mathcal{A}_r(m^2_-,
m^2_+) + a_\textrm{NR}e^{i\phi_\textrm{NR}}
\nonumber \\
\overline{\mathcal{A}} 
(m^2_-,m^2_+) = \sum_r \overline{a}_r e^{i\overline{\phi}_r} 
\mathcal{A}_r(m^2_-,
m^2_+) + \overline{a}_\textrm{NR}e^{i\overline{\phi}_\textrm{NR}}
\nonumber 
\end{eqnarray}
where all resonant amplitudes ($\mathcal{A}_r$)
are summed up with relative phase
information as well as non-resonant (NR) terms. 
From above, time-dependent decay rate parameters such
as 
$e^{-\Gamma t} \cos{x\Gamma t}$,
$e^{-\Gamma t} \sin{x\Gamma t}$,
and 
$e^{[-(1\pm y)\Gamma t]}$
can be extraced from the measurements. This is a fairly complicated
analysis since it contains 18 quasi two-body Dalitz-plot parameters
with time-dependent unbinned maximimum likelihood analysis. Both
Belle and BaBar collaborations analyze their data. 
BaBar has measured~\cite{ref:babar_dalitz}
$x=(0.16\pm0.23\pm0.14)\%$,
and
$y=(0.57\pm0.20\pm0.15)\%$.
On the othe hand, Belle has measured $|q/p|$ and $\phi$ in addition to $x$ and
$y$~\cite{ref:belle_dalitz}:
$x=(0.80\pm0.29^{+0.13}_{-0.16})\%$,
$y=(0.33\pm0.24^{+0.10}_{-0.14})\%$,
$|q/p|=0.86\pm0.30\pm0.09$,
and
$\phi=-0.24\pm0.30\pm0.09$. Figure \ref{fig:belle_dalitz} shows
the allowed region measured by the Belle collaboration~\cite{ref:belle_dalitz}.

\begin{figure}
\includegraphics[width=80mm]{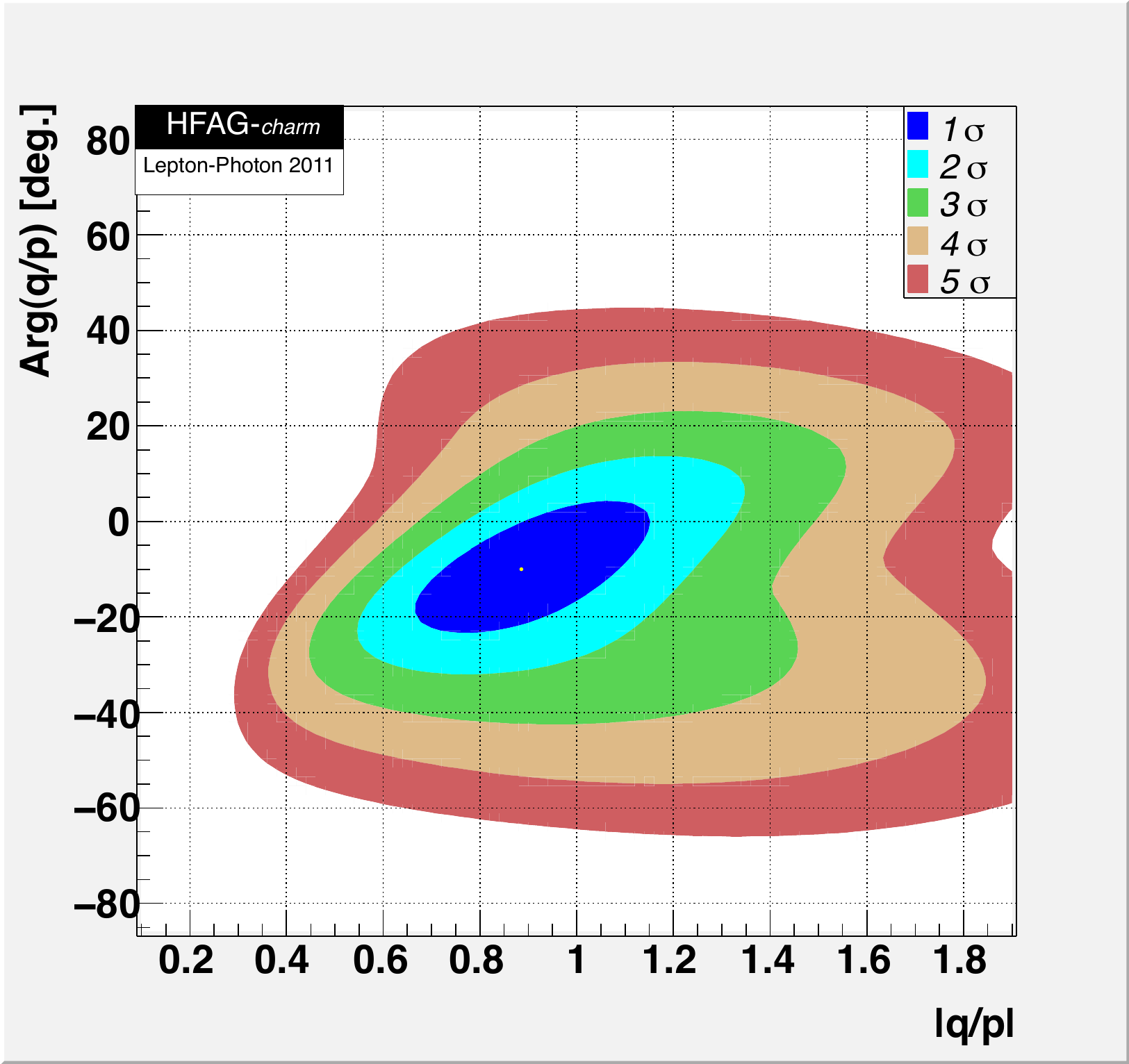}
\caption{
The world average contour of the mixing parameters $x$ and $y$, indicating
non-zero values of $x$ and $y$~\cite{ref:hfag}.
\label{fig:hfag_1}
}
\end{figure}

\begin{figure}
\includegraphics[width=80mm]{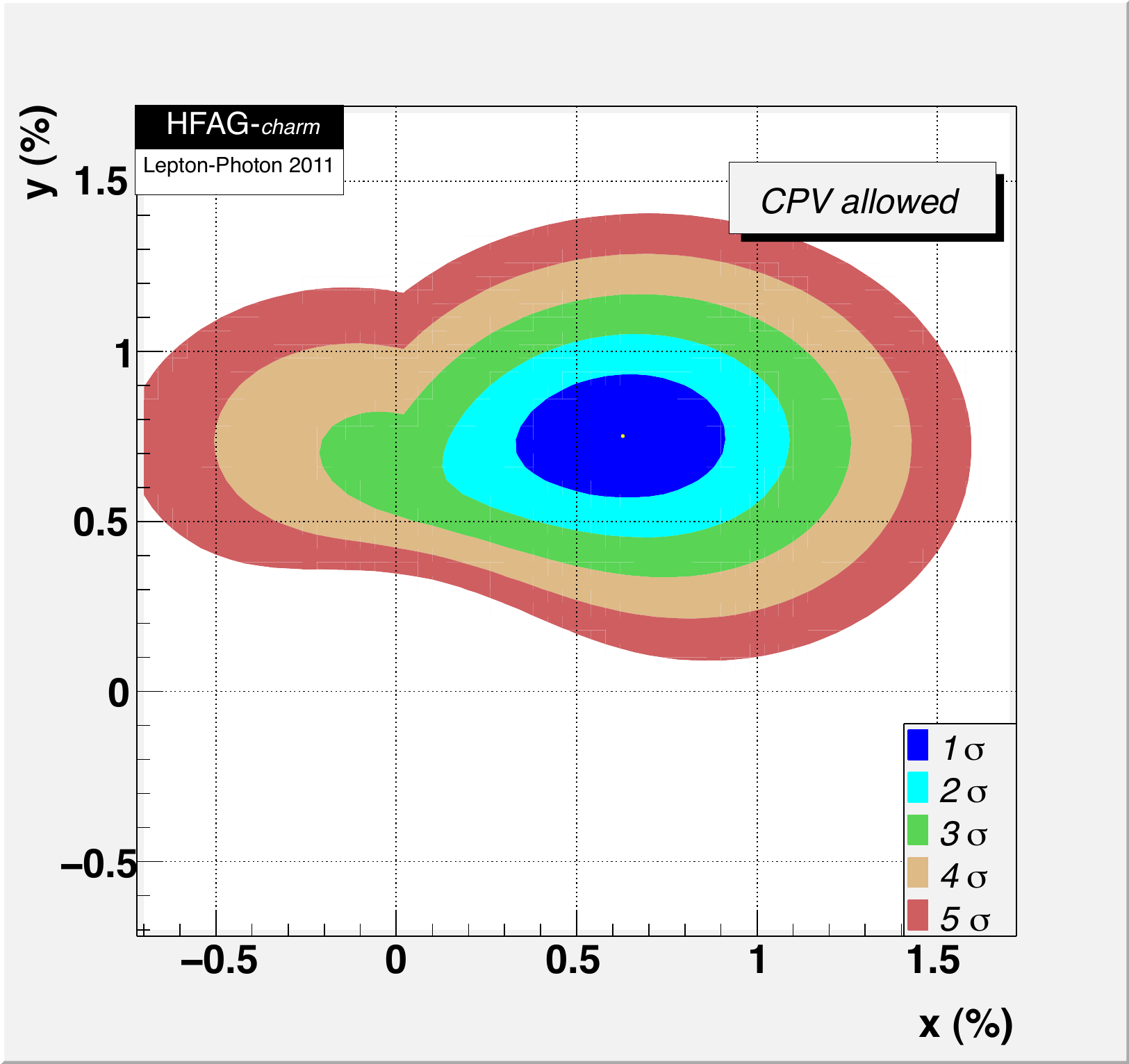}
\caption{
The world average contour of the $CP$ violation parameters in the mixing.
Vertical axis is for $\phi$ and the horizontal axis is for
$|q/p|$~\cite{ref:hfag}.
\label{fig:hfag_2}
}
\end{figure}

\subsection{World Average of Mixing Parameters}

 The heavy flavor averaging group~\cite{ref:hfag} average
of $D^0$ mixing/$CP$ violation underlying physics parameters from
existing observables are summarized in Fig.~\ref{fig:hfag_1}
and Fig.~\ref{fig:hfag_2}. Even if the impact is not significant,
the recent results from the LHCb experiment are included here~\cite{ref:lhcb}.
The world average values are 
$x=(0.63^{+0.19}_{-0.20})$\%,
$y=(0.75^\pm 0.12)$\%,
$|q/p|=0.89^{+0.17}_{-0.15}$, and 
$\phi(^\circ)=-10.0^{+9.4}_{-8.8}$. 
These indicate that no mixing senario is excluded at 10$\sigma$ level but
there is no indication of $CP$ violation in the $D^0$ mixing.
From this, one can test various NP models~\cite{ref:golowich}.
For example, from the present bound of $x$, one can constrain 
the 4th generation quark doublet coupling as 
$|V_{ub^{\prime}}V_{cb^{\prime}}|<10^{-3}$ for $m_{b^{\prime}}$ = 500 GeV.

\section{$CP$ Violation in $D$ Meson Decays}

 For the time-integrated search for $CP$ violation, one has to
extract the $CP$ asymmetry from detector effect and production asymmetry
in the observed asymmetry. To a good approximation,
\begin{eqnarray}
A_\textrm{rec} = 
\frac{
N^D_\textrm{rec} - N^{\overline{D}}_\textrm{rec}
}{
N^D_\textrm{rec} + N^{\overline{D}}_\textrm{rec}
}
\cong
A_{CP} + A_{FB} + A_\epsilon
\end{eqnarray}
where $N^{D}$ and $N^{\overline{D}}$ are reconstructed yields for
$D$ and $\overline{D}$ mesons, respectively.
$A_\textrm{rec}$ is the reconstructed asymmetry, $A_{CP}$ is the
$CP$ asymmetry, $A_{FB}$ is the forward-backward asymmetry in
the production, and $A_\epsilon$ is the one due to the charged
particle reconstruction efficiency asymmetry.
To remove asymmetries that do not originate from $CP$ violation,
various techniques have been developed depending on availablity
of control samples of particular decay mode of interest. The important
point here is that to control systematics, one really has to use real
data.

\subsection{Time-integrated Search for $CP$ Violation}

 The CDF collaboration
analyzes the data of $D^0 \rightarrow K^+K^-$ and $D^0 \rightarrow
\pi^+\pi^-$~\cite{ref:cdf_cp}. The formulation of various asymmetries
in this case is follows:
\begin{eqnarray}
A_\textrm{rec}(hh^*) &=& A_{CP}(hh) + A_\epsilon(\pi_s)^{hh^{*}}
\nonumber \\
A_\textrm{rec}(K\pi^*) &=& A_{CP}(K\pi) + A_\epsilon(\pi_s)^{K\pi^{*}}
+ A_\epsilon(K\pi)^{K\pi^{*}}
\nonumber \\
A_\textrm{rec}(K\pi) &=& A_{CP}(K\pi) ~~~~~~~~~~~~~~~~~+ A_\epsilon(K\pi)^{K\pi}
\nonumber \\
A_{CP}(hh) &=& 
A_\textrm{rec}(hh^*) - A_\textrm{rec}(K\pi^*) + A_\textrm{rec}(K\pi)
\nonumber 
\end{eqnarray}
where (*) indicates that the $D^0$ meson flavor is tagged with slow pions.
Note that the last equation is obtained by computing a linear combnation of
above three. Using the relation above, 
$A_{CP}(\pi\pi) = (+0.22\pm0.24\pm0.11)$\% 
and
$A_{CP}(KK) = (-0.24\pm0.22\pm0.10)$\% 
are obtained. Both direct ($a^\textrm{dir}_{CP}$) and mixing-induced $CP$ 
violation ($a^\textrm{ind}_{CP}$) contribute to the asymmetry as
$A_{CP} = a^\textrm{ind}_{CP} + \frac{\langle t \rangle}{\tau} 
a^\textrm{ind}_{CP}$ where $\tau$ is the mean lifetime of $D^0$ meson
and $\langle t \rangle$ is the mean value of decay time distribution
from the measurement. Note that the measurement errors are smaller than
those from $B$-factories.

\begin{figure}
\includegraphics[width=85mm]{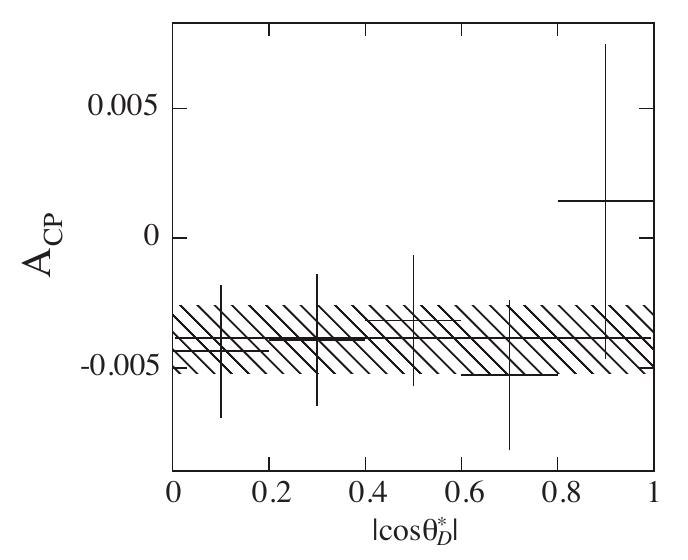}
\caption{
$CP$ asymmetries for $D^+ \rightarrow K^0_S \pi^+$ candidates  as  a 
function of $|\cos{\theta^*_D}|$ in the data sample. The solid line represents 
the central value of $A_CP$ and the hatched region is the 
$\pm1\sigma$ interval , obtained from a minimization assuming no 
dependence on $|\cos{\theta^*_D}|$.
\label{fig:babar_ksp}
}
\end{figure}

 Time-integrated $CP$ violation search in $D^+ \rightarrow K^0_S \pi^+$
is carried out by both the Belle and the BaBar 
collaborations~\cite{ref:belle_ksp,ref:babar_ksp}. In both experiments,
one has to remove the forward-backward asymmetry from the production and  
the asymmetry from the charged particle reconstruction. To correct for 
them, Belle
uses $D^+_s \rightarrow \phi \pi^+$ and $D^0 \rightarrow K^- \pi^+$ decay
modes. The measured asymmetry is $A_{CP}(D^+ \rightarrow K^0_S \pi^+)
=(-0.71\pm0.19\pm0.20)$\% where the major systematical uncertainty is from
the statistics of $D^+_s \rightarrow \phi \pi^+$ sample (0.18\%). On the
other hand, BaBar uses inclusive data of on- and off-resonance data for 
the correction and this enables one to reduce the systematical uncertainty
due to the correction down to 0.08\%. The corrected asymmetry value
from BaBar is $A_{CP}(D^+ \rightarrow K^0_S \pi^+)=(-0.44\pm0.13\pm0.10)$\%.   
The $CP$ asymmetry from BaBar is shown in Fig.~\ref{fig:babar_ksp} as a function
of $|\cos{\theta^*_D}|$ indicating a weak deviation from zero. If results
from two experiments are combined, one gets 
$A_{CP}(D^+ \rightarrow K^0_S \pi^+)=(-0.51\pm0.14)$\% and this may be
the first hint of $CP$ violation in the charm sector. On the other hand,
this is consistent with $CP$ violation from neutral kaon mixing 
$(-0.332\pm0.006)$\%~\cite{ref:pdg2010}.

 $A_{CP}(D^0\rightarrow K^0_S P^0)$, 
where $P^0$ is $\pi^0$ or $\eta^{(\prime)}$ 
is measured by the Belle experiment~\cite{ref:belle_p0}. 
The decay $D^{*+} \rightarrow 
D^0\pi^+_s$ is used in order to identify the flavor of the $D^0$ meson,
and to correct for $A^{\pi^{+}_{s}}$, $D^0 \rightarrow K^- \pi^+$ (untagged)
and 
$D^{*+} \rightarrow D^0 \pi^+_s \rightarrow K^-\pi^+ \pi^+_s$ (tagged)
are used. The results are 
$A_{CP}(D^0 \rightarrow K^0_S \pi^0)=(-0.28\pm0.19\pm0.10)$\%,
$A_{CP}(D^0 \rightarrow K^0_S \eta)=(+0.54\pm0.51\pm0.16)$\%,
and
$A_{CP}(D^0 \rightarrow K^0_S \eta^\prime)=(+0.98\pm0.67\pm0.14)$\%.
One can assume $A_{CP}(D^0 \rightarrow K^0_S \pi^0) = A^{K^0}_{CP} 
+ a^\textrm{ind}$ and therefore can extract $a^\textrm{ind}$. They
extract $a^\textrm{ind}(D^0\rightarrow K^0_S \pi^0)=(+0.05\pm0.19\pm0.10)$\%
and it can be compared with the previous measurement done by the 
Belle experiment~\cite{ref:belle_mix}: 
$a^\textrm{ind}(D^0\rightarrow K^+ K^-)=(-0.01\pm0.30\pm0.15)$\% and 
this is the first experimental test of the universality of $a^\textrm{ind}$
in $D^0$ decays.

\begin{figure}
\includegraphics[width=0.50\textwidth]{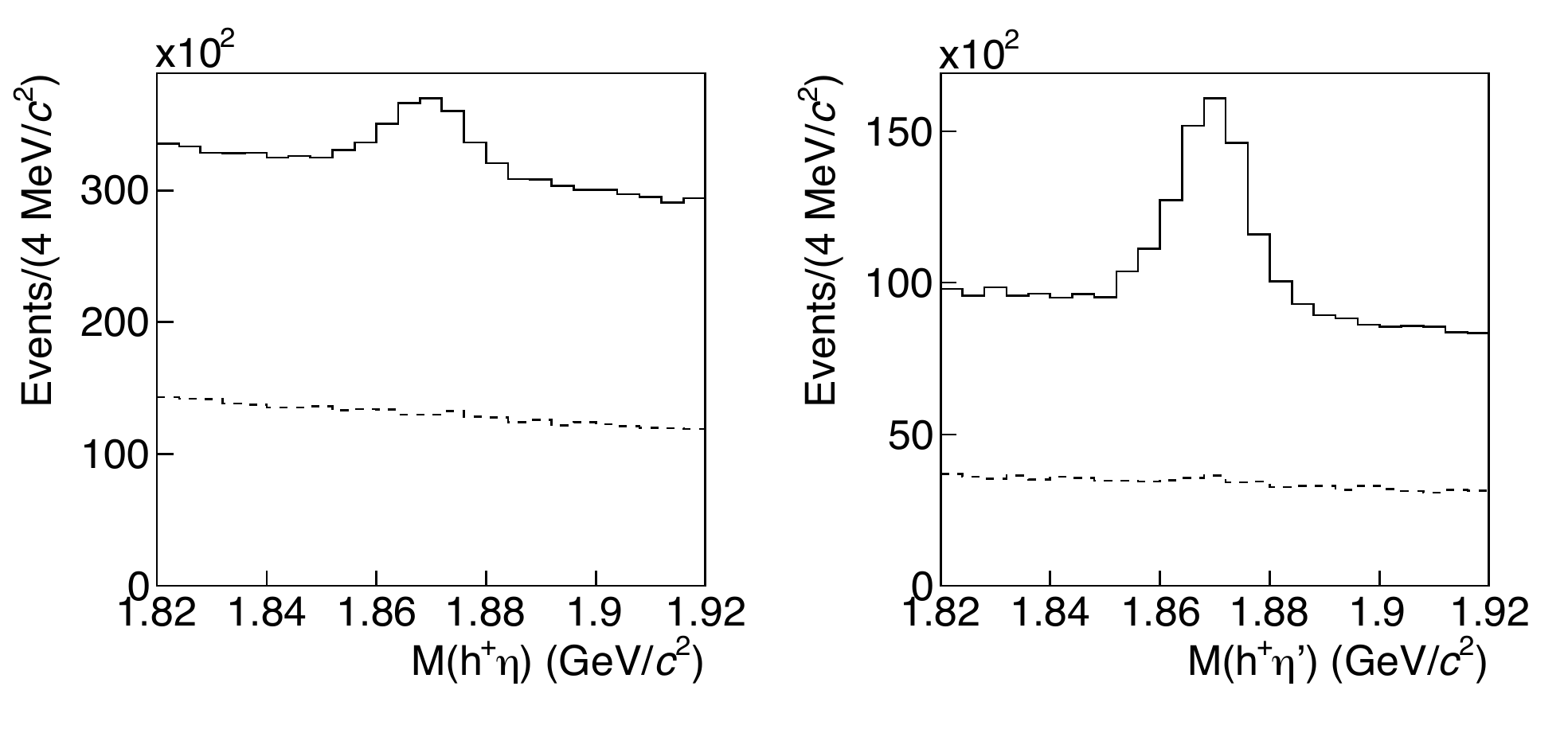} \quad
\caption{
The invariant mass distributions of $h^+ \eta$
($h^+ \eta'$) in the left (right) plot after the initial selection.
The solid histograms show $\pi^+ \eta^{(\prime)}$ while the dashed histograms
show  $K^+ \eta^{(\prime)}$ final states~\cite{ref:belle_eta}.
\label{fig:belle_eta_1}
}
\end{figure}

\begin{figure}
\mbox{
\includegraphics[width=0.22\textwidth]{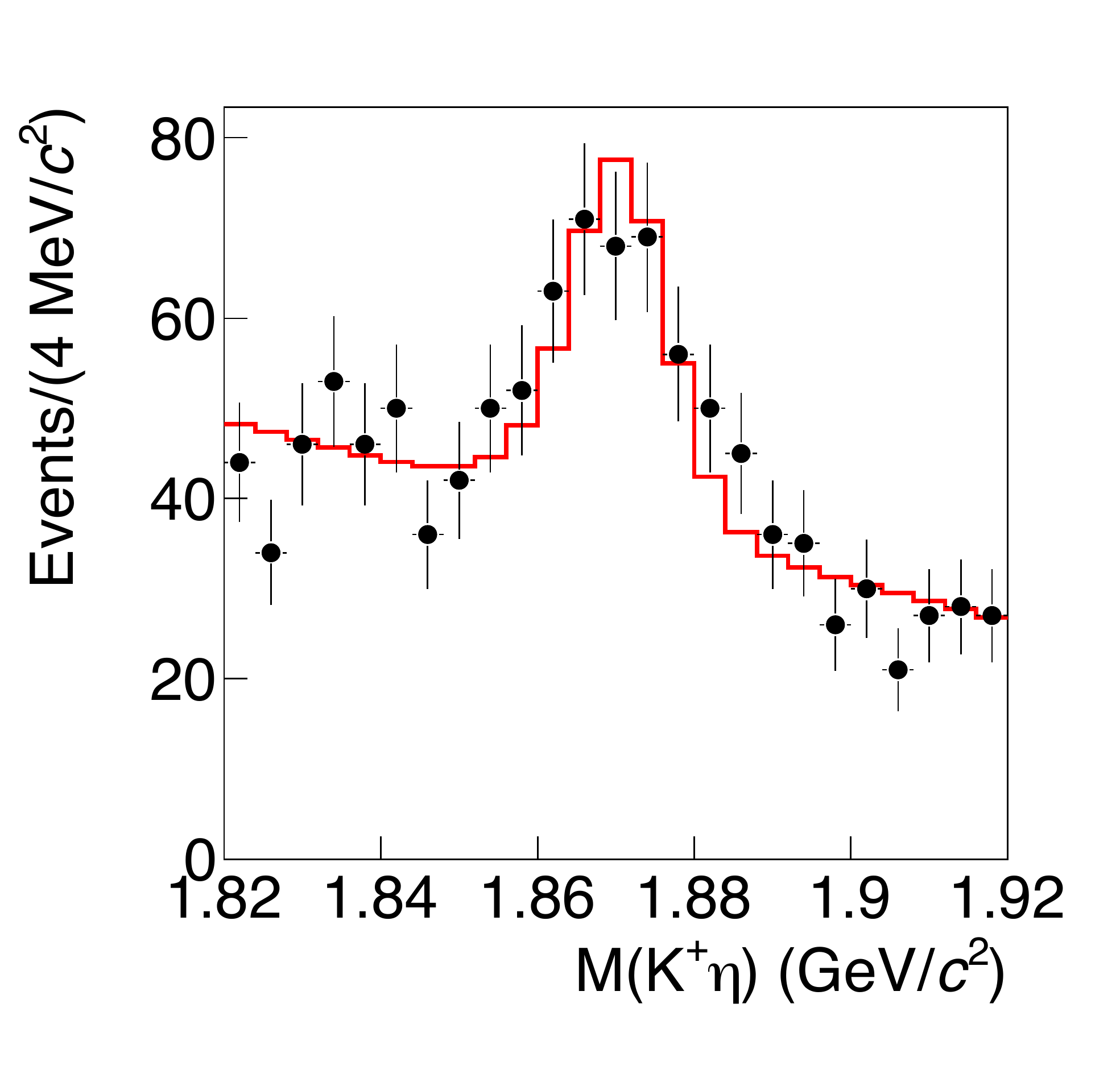} \quad
\includegraphics[width=0.22\textwidth]{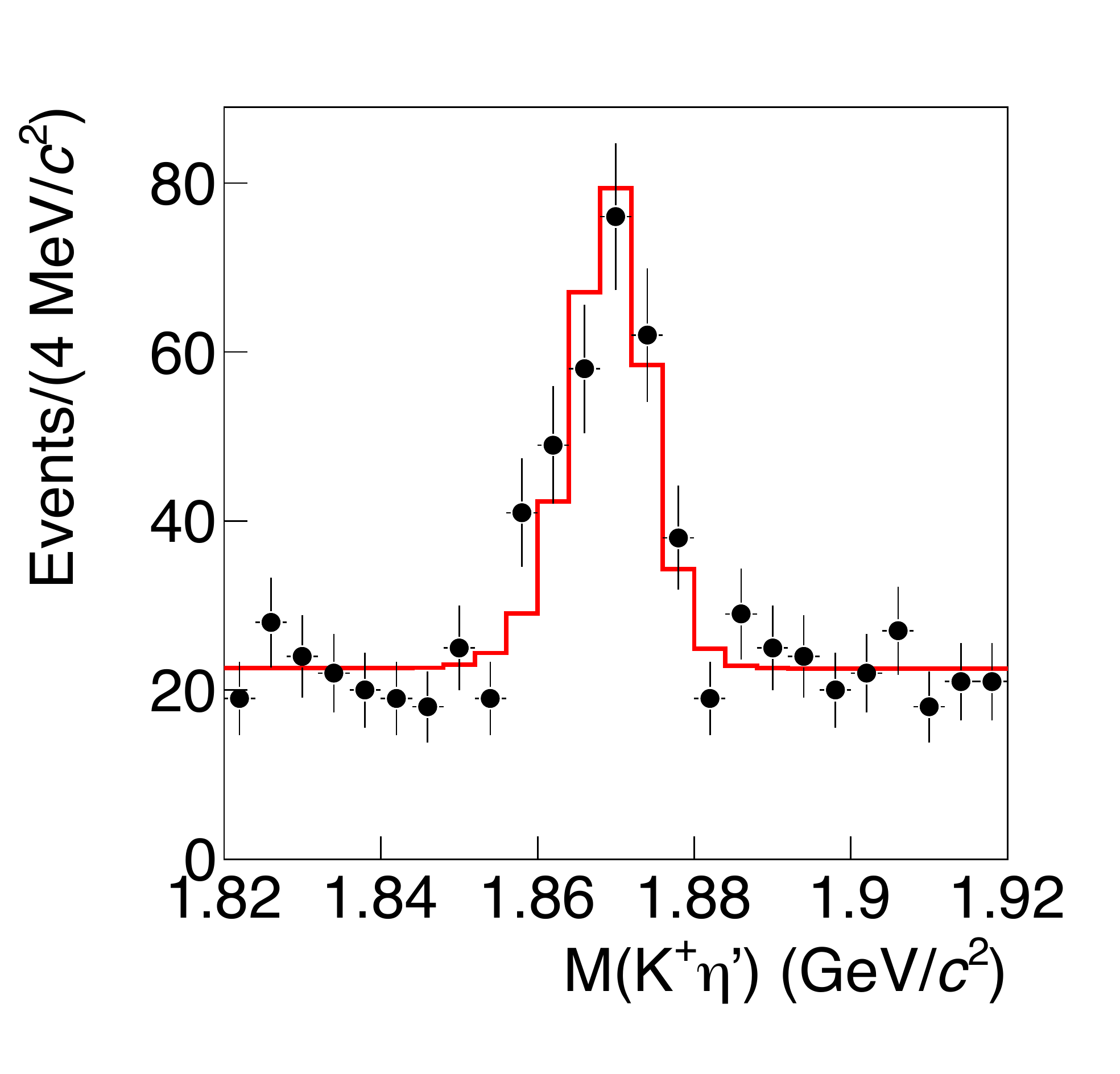}
}
\caption{
The invariant mass distributions used for branching fraction measurements
of $D^+ \rightarrow K^+ \eta$ (left)
and
of $D^+ \rightarrow K^+ \eta^\prime$ (right).
Points with error bars and histograms correspond to the data and
the fit, respectively~\cite{ref:belle_eta}.
\label{fig:belle_eta_2}
}
\end{figure}

 The first observations of DCSD modes $D^+ \rightarrow K^+ \eta^{(\prime)}$
is made by the Belle experiment~\cite{ref:belle_eta}. 
When standard criteria are imposed, there is little indication of
the signal as shown in Fig.~\ref{fig:belle_eta_1}.
In order to extract
the such small signals from the backgrounds, extremely tight selection
criteria are imposed based on a grid search technique~\cite{ref:grid}.
Based on our grid search, a tight set of selections is imposed to the
data and clear signals are first observed as shown in 
Fig.~\ref{fig:belle_eta_2}.
From these, relative branching fractions are extracted to be
$\mathcal{B}(D^+ \rightarrow K^+ \eta)$/$\mathcal{B}(D^+ \rightarrow \pi^+ \eta)$
= (3.06 $\pm$ 0.43 $\pm$ 0.14)\%
and
$\mathcal{B}(D^+ \rightarrow K^+ \eta')$/$\mathcal{B}(D^+ \rightarrow \pi^+ \eta')$
= (3.77 $\pm$ 0.39 $\pm$ 0.10)\%.
Using the relations in Ref.~\cite{ref:dcsd_th}, which give
\begin{eqnarray}
|T|^2 &=& 3|\mathcal{A}(K^+\eta)|^2
\nonumber \\
|A|^2 &=& \frac{1}{2}\Bigg[|\mathcal{A}(K^+\pi^0)|^2
+|\mathcal{A}(K^+ \eta^\prime)|^2\Bigg] 
\nonumber \\
&-&
|\mathcal{A}(K^+\eta)|^2
\nonumber \\
\cos{\delta_{TA}} &=& \frac{1}{2|T||A|}
\Bigg[
2|\mathcal{A}(K^+\eta)|^2 + \frac{1}{2}|\mathcal{A}(K^+\eta^\prime)|^2 
\nonumber \\
&-& \frac{3}{2}|\mathcal{A}(K^+ \pi^0)|^2
\Bigg]
\end{eqnarray}
where $T$ ($A$) is the tree (annihilation) amplitude
and $\mathcal{A}$ is the specified decay amplitude, and
from the recent branching fraction measurement of $\mathcal{B}(D^+ \rightarrow
K^+ \pi^0) = (1.72\pm 0.20) \times 10^{-4} $~\cite{ref:cleo}, they find that
the relative final-state phase difference between the
tree and annihilation in $D^+$ decays,
$\delta_{TA}$, is (72 $\pm$ 9)$^\circ$ or
(288 $\pm$ 9)$^\circ$. This is the first experimenetal access to the
phase difference between the tree and annihilation amplitudes 
in these decay modes.

\subsection{Time Reversal Violation}

 Under the assumption of $CPT$ invariance, $T$-violation is a signal
for $CP$ violation and therefore it can be studied via $T$-violation
search. For a multi-particle ($>3$) final state, one can form a kinematic
product that is odd under time reveral as
\begin{eqnarray}
C_T = \mathbf{p}_1 \cdot (\mathbf{p}_2 \times \mathbf{p}_3)
\end{eqnarray}
where $\mathbf{p}_i$ is the momemtum vector of a daughter particle $i$. Note
that at least four different particles are required in the final state
so that three of them are independent. The strong interaction dynamics
can produce a non-zero value of the asymmetries:
\begin{eqnarray}
A_T &\equiv& \frac{\Gamma(C_T>0)-\Gamma(C_T<0)}
{\Gamma(C_T>0)+\Gamma(C_T<0)}
\nonumber \\
\overline{A}_T &\equiv& \frac{\Gamma(-\overline{C}_T>0)-\Gamma(-\overline{C}_T<0)}
{\Gamma(-\overline{C}_T>0)+\Gamma(-\overline{C}_T<0)}
\end{eqnarray}
where the second equation is the asymmetry for the $CP$-conjugate decay 
process. However, the difference characterizes $T$-violation in the
weak decay process as
\begin{eqnarray}
\mathcal{A}_T = \equiv \frac{1}{2} (A_T - \overline{A}_T).
\end{eqnarray}
This $T$-violation observable is measured by the BaBar experiment
in decays of $D^0 \rightarrow K^+K^- \pi^+ \pi^-$ and
$D^+_{(s)} \rightarrow K^+ K^0_S \pi^+ \pi^-$~\cite{ref:babar_t}.
No evidence of $T$-violation is seen and results are
$\mathcal{A}_T(D^0\rightarrow K^+K^-\pi^+\pi^-)=(1.0\pm5.1\pm4.4)\times 10^{-3}$,
$\mathcal{A}_T(D^+\rightarrow K^+K^0_S\pi^+\pi^-)=(-12.0\pm10.1\pm4.6)\times 10^{-3}$,
and
$\mathcal{A}_T(D^+_s\rightarrow K^+K^0_S\pi^+\pi^-)=(-13.6\pm7.7\pm3.4)\times 10^{-3}$.

\begin{table}[t]
\begin{center}
\caption{Expected sensitivity with 5 ab$^{-1}$ and with 50 ab$^{-1}$
in future Belle-II experiment.}
\begin{tabular}{|c|c|c|c|}
\hline \textbf{Parameters} & \textbf{Present} & \textbf{5 ab$^{-1}$} &
\textbf{50 ab$^{-1}$}
\\
 & \textbf{uncertainty} & & \\
\hline 
$y_{cp}$  &     $\pm$0.39 & $\pm$0.12 & $\pm$0.05 \\
$A_{\Gamma}$  & $\pm$0.33 & $\pm$0.10 & $\pm$0.04 \\
$x$(\%)       & $\pm$0.31 & $\pm$0.10 & $\pm$0.03 \\
$y$(\%)       & $\pm$0.26 & $\pm$0.08 & $\pm$0.03 \\
$|q/p|$       & $\pm$0.30 & $\pm$0.10 & $\pm$0.03 \\
$\phi$(rad)   & $\pm$0.30 & $\pm$0.10 & $\pm$0.03 \\
\hline
\end{tabular}
\label{table:belle2}
\end{center}
\end{table}

\section{Summary and Outlook}

 We reviewed present status of $D^0$ mixing/$CPV$ and $D$ decays.
The mixing, the oscillation of $D^0$ meson flavor, has been firmly
established~\cite{ref:hfag}, but not from a single experiment yet.
The search for $CP$ violation in $D$ meson decays are carried out 
and show no $CP$ violation effect down to $\mathcal{O}(10^{-3})$.
There may be a hint of $CP$ violation 
in $D^+ \rightarrow K^0_S \pi^+$ decay mode but it is consistent
with $CP$ violation from the kaon mixing. 

 However, we are entering a new era of flavor physics, in particular
for the charmed meson physics. 
The remaining yet to be analyzed data from
present $B$-factories will produce even higher-sensitive results
in near future.
The LHCb is starting to demonstrate
great ability to reconstruct charmed mesons with extremely low
background~\cite{ref:lhcb}. Also, planned super $B$-factories
are expected to reach highest amount of data sample ever achieved,
to 50 ab$^{-1}$~\cite{ref:belle2}. Table~\ref{table:belle2} lists
expected sensitivity for various mixing and $CP$ violation parameters
for the Belle-II experiment, expecting to explore physics beyond SM,
mostly through quantum loops in the reaction.

\bigskip 
\begin{acknowledgments}
We thank the KEKB group for excellent operation of the
accelerator, the KEK cryogenics group for efficient solenoid
operations, and the KEK computer group and
the NII for valuable computing and SINET4 network support.
We acknowledge support from MEXT, JSPS and Nagoya's TLPRC (Japan);
ARC and DIISR (Australia); NSFC (China); MSMT (Czechia);
DST (India); MEST, NRF, NSDC of KISTI, and WCU (Korea); MNiSW (Poland);
MES and RFAAE (Russia); ARRS (Slovenia); SNSF (Switzerland);
NSC and MOE (Taiwan); and DOE (USA). E. Won
acknowledges support by NRF Grant No. 2011-0027652 and B. R. Ko
acknowledges support by NRF Grant No. 2011-0025750.
\end{acknowledgments}

\bigskip 
\bibliography{basename of .bib file}

\end{document}